\documentclass[structabstract]{aa}
\usepackage{graphicx}
\usepackage{txfonts}
\usepackage{natbib}
\begin{document}
   \title{The kinematics of the quadrupolar nebula M~1--75 and the identification of its central star\thanks{Based on
observations made with the 4.2~m William Herschel Telescope and the 2.5~m Isaac Newton Telescope, both operated on the island of la Palma by the Isaac Newton Group of Telescopes in the Spanish Observatorio del Roque de los Muchachos of the Instituto de Astrof\'\i sica de Canarias.}}


\author{M. Santander-Garc\'\i a\inst{1,2,3}
    \and  P. Rodr\'iguez-Gil\inst{1,2,3}
    \and O. Hernandez\inst{4}
    \and R. L. M. Corradi\inst{2,3}
    \and D. Jones\inst{1,5}
    \and C. Giammanco\inst{2,3}
    \and J. E. Beckman\inst{6,2,3}
    \and C. Carignan\inst{4}
    \and K. Fathi\inst{7}
    \and M. M. Rubio-D\'\i ez\inst{1,8}
    \and  F. Jim\'enez-Luj\'an\inst{1,9,10}
    \and  C. R. Benn\inst{1}
            }

\institute{
         Isaac Newton Group of Telescopes, Ap.\ de Correos 321, E-38700 Sta. Cruz de la Palma, Spain
         \\e-mail: msantander@ing.iac.es    
        \and
         Instituto de Astrof\'\i sica de Canarias, E-38200 La Laguna, Tenerife, Spain
        \and
         Departamento de Astrof\'\i sica, Universidad de La Laguna, E-38205 La Laguna, Tenerife, Spain
	\and
	    LAE, Universit\'e de Montr\'eal, C.P. 6128 Succ. Centre Ville, Montr\'eal, QC H3C 3J7, Canada
    \and
	    Jodrell Bank Centre for Astrophysics, School of Physics and Astronomy, University of Manchester, Manchester, M13 9PL, UK
         \and
              Consejo Superior de Investigaciones Cient\'\i ficas, Madrid, Spain
         \and
              Stockholm Observatory, Department of Astronomy, Stockholm University, AlbaNova, 106 91 Stockholm, Sweden 
	\and
	    Centro de Astrobiolog\'\i a, CSIC-INTA, Ctra de Torrej\'on a Ajalvir km 4, E-28850 Torrej\'on de Ardoz, Spain
	\and
	    Instituto de F\'\i sica de Cantabria (CSIC-Universidad de Cantabria), E-39005 Santander, Cantabria, Spain
	\and
	Dpto. de F\'isica Moderna, Universidad de Cantabria, Avda de los Castros s/n,  E-39005 Santander, Cantabria, Spain
        }

   \date{A\&A: Received April 21, 2010; accepted June 10, 2010}


  \abstract
   {The link between the shaping of bipolar planetary nebulae and their central stars is still poorly understood.}
   {The kinematics and shaping of the multipolar nebula M~1--75 are hereby investigated, and the location and nature of its central star are briefly discussed.}
   {Fabry-Perot data from GH$\alpha$FAS on the WHT sampling the Doppler shift of the [N{\sc ii}] 658.3 nm line are used to study the dynamics of the nebula, by means of a detailed 3-D spatio-kinematical model. Multi-wavelength images and spectra from the WFC and IDS on the INT, and from ACAM on the WHT, allowed us to constrain the parameters of the central star.}
   {The two pairs of lobes, angularly separated by $\sim$22$^\circ$, were ejected simultaneously approx. $\sim$3500-5000 years ago, at the adopted distance range from 3.5 to 5.0~kpc. The larger lobes show evidence of a slight degree of point symmetry. The shaping of the nebula could be explained by wind interaction in a system consisting of a post-AGB star surrounded by a disc warped by radiative instabilities. This requires the system to be a close binary or a single star which engulfed a planet as it died.  On the other hand, we present broad- and narrow-band images and a low S/N optical spectrum of the highly-reddened, previously unnoticed star which is likely the nebular progenitor. Its estimated $V-I$ colour allows us to derive a rough estimate of the parameters and nature of the central star.}
   {}


   \maketitle
%

\section{Introduction}

Planetary nebulae (PNe) represent the terminal breath of 90$\%$ of the stars in the Universe. However, their shaping mechanism is still poorly understood.

Bipolar PNe are undoubtedly the most challenging case. Several attempts have been made to explain their shaping (see the review by \citealp{balick02}), breaking spherical symmetry by invoking elements which fall in two distinct categories: {\it a)} rapid stellar rotation and/or magnetic fields \citep[e.g.][]{garciasegura99,blackman01a}, and {\it b)} a close interacting companion to the star (e.g. \citealt{nordhaus06}, for a review see \citealt{demarco09}). This latter hypothesis seems to be gaining some ground as close binary systems are progressively being found (e.g. \citealp{miszalski09a}; \citealp{miszalski10}) at the cores of bipolar PNe.

Spatio-kinematical modelling of PNe constitutes an excellent tool to test theoretical models. It provides us with important parameters to be matched by the different models of formation, such as the  3-D morphologies and velocity fields of the outflows, their kinematical age (once disentangled from the distance to the nebula) and their orientation to the line of sight.

M~1--75 (PN G068.8-00.0, $\alpha$ = 20 04 44.086 $\delta$ = +31 27 24.42 J2000) is a good example of a complex nebula. It displays a seemingly irregular horseshoe-like central region, out of which two systems of faint lobes emerge. It was first classified as quadrupolar by \citet{manchado96b}, and a tentative attempt to recover its kinematic parameters was done by \citet{dobrincic08}.

In this paper we present Fabry-Perot interferometry of \mbox{M~1--75}, from which we derive a detailed spatio-kinematical model (section 3). We also report the first imaging and spectroscopic detection of its central star (section 4). We then discuss both results and their implications in the shaping of the nebula in section 5.


\section{Observations and Data reduction}

\subsection{Fabry-Perot interferometric data}

The [N{\sc ii}] 658.3 nm emission of M~1-75 was scanned with GH$\alpha$FAS (Galaxy H$\alpha$ Fabry-Perot System) on the 4.2~m WHT (William Herschel Telescope) on July 6, 2007, as part of its commissioning programme. The nebula was observed in high-resolution mode with the OM4 etalon (resolving power R$\sim$18000, effective finesse $\Im_\mathrm{e}$=24) and a plate scale of 0$''$.2 pixel$^{-1}$. The free spectral range was 8.62 $\AA$ or 392~km~s$^{-1}$ split into 48 channels, thus leading to a velocity step of 8.16 km~s$^{-1}$ per channel. The total exposure time of the scanning was 1.9~hr, and the seeing 0$''$.8. The instrumental response function (IRF) was measured by fitting a Lorentzian to the profile of a Neon lamp line and resulted in an instrumental width (FWHM) of 18.6~km~s$^{-1}$.

\begin{figure}
\center
\resizebox{7.5cm}{!}{\includegraphics{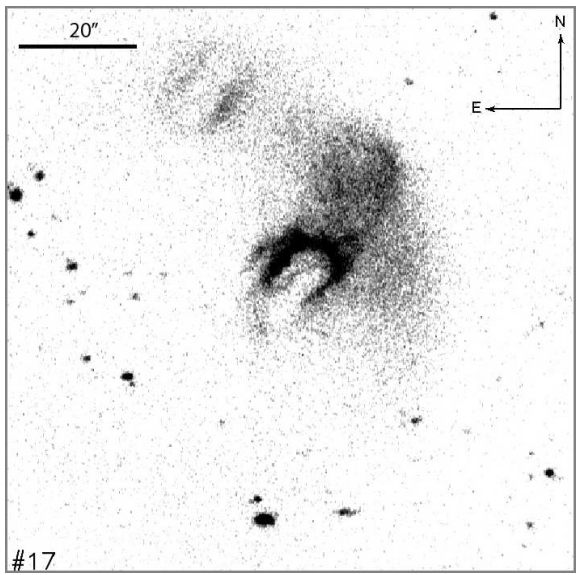}}
\caption{Channel \#17 of the GHAFAS datacube, showing the blue-shifted upper side of the horseshoe (centre). The faint emission from the large lobe extending northwest is contaminated by a broad arc-shaped artifact which spans over several channels. A fainter version of the horseshoe itself is replicated as a ghost near the top left corner of the frame. Additionally, several channels are slightly contaminated by H$\alpha$ emission from adjacent orders. These artifacts, however, do not prevent a proper modelling of the nebula (see section 3).}
\label{F1}
\end{figure}

The data were reduced following the standard procedure for GH$\alpha$FAS data, which are described in \citet{hernandez08}. Several artifacts persisted through the data reduction process. These include slight contamination by H$\alpha$ emission from adjacent orders (specially in the first and last channels of the datacube), a ghost of the inner region of the nebula, and an arc-shaped artifact which runs across several channels, at different locations (see Fig.~\ref{F1}). 

\subsection{Broad and narrow-band imaging}

Several images of M~1--75 in the light of different filters (U, B, V, I, H$\alpha$ and Str\"omgren Y) were taken with ACAM (Auxiliary-port Camera) on the WHT and with the WFC (Wide Field Camera) on the 2.5~m INT (Isaac Newton Telescope). The log of the observations can be found in Table 1.

All these data were reduced following standard {\tt IRAF}\footnote{{\tt IRAF} is distributed by National Optical Astronomy Observatories.} procedures.

\begin{table}[!h] 
\begin{center} 
\begin{tabular}[t]{lccccc}
\hline\noalign{\smallskip}
Date & Telescope/  &  filter & Band & Exp. time   & seeing   \\    
         &  Instrument  &    ref.  &           & (s)               &        \\    
\hline\noalign{\smallskip}
2009 Jun 11   &   WHT/ACAM   &   \#17   &   I                             &   3$\times$120    &   1$''$.4 \\
2009 Jun 12   &   INT/WFC        &   \#201   &   Str. Y   &  2$\times$600     &   1$''$.3 \\
2009 Jun 12   &   INT/WFC        &   \#197   &   H$\alpha$         &  120                       &   1$''$.3 \\
2009 Sep 10   &   INT/WFC        &   \#204   &   U                        &  1200                     &   1$''$.6 \\
2009 Sep 10   &   INT/WFC        &   \#204   &   B                        &  1200                     &   1$''$.6 \\
2009 Sep 10   &   INT/WFC        &   \#204   &   V                        &  1800                     &   1$''$.6 \\
2009 Sep 10   &   INT/WFC        &   \#204   &   I                          &  600                       &   1$''$.6 \\
2009 Sep 10   &   INT/WFC        &   \#204   &   Str. Y  &  600                       &   1$''$.6 \\
\noalign{\smallskip}\hline
\end{tabular}
\end{center} 
\label{T1} \caption{Log of the broad and narrow-band imaging observations.} 
\end{table}

\subsection{Long-slit spectroscopy}

An 3600~s spectrum with the slit at parallactic angle (P.A.=284$^\circ$), crossing the centre of the inner nebula,  was taken with IDS (Intermediate Dispersion Spectrograph) on the INT on March 9, 2009. The R300V grating was used, centered at 540 nm and effectively covering from 430 to 810 nm at a resolving power R$\sim$1500. The slit width was 1$''$, while the seeing was 1$''$.8. HD~192281 was chosen as the standard star to account for flux and sensitivity calibration.

A low-resolution (resolving power R ranging from 290 and 570), 40~min spectrum of M1-75 was taken with ACAM on the WHT on June 11, 2009, with the 400 lines~mm$^{-1}$ transmission VPH (Volume Phase Holographic) disperser, covering the wavelength range between 350 and 950 nm. The 1$''$ wide slit  was positioned at P.A.=0$^\circ$ in order to get the light from the two central star candidates (see section 4). The seeing was 2$''$.8, and the standard star was HD~338808.

The spectra were de-biased, flat-fielded, distortion-corrected and  wavelength calibrated (from copper-argon and copper-neon arc lamps) using standard {\tt IRAF} routines. After extraction of the selected nebular and central star features from the orthogonal 2-D spectra, the 1-D spectra were telluric and sensitivity corrected using the spectrum of the spectrophotometric standard star.


\section{An improved spatio-kinematical model}

\begin{figure*}
\center
\resizebox{15cm}{!}{\includegraphics{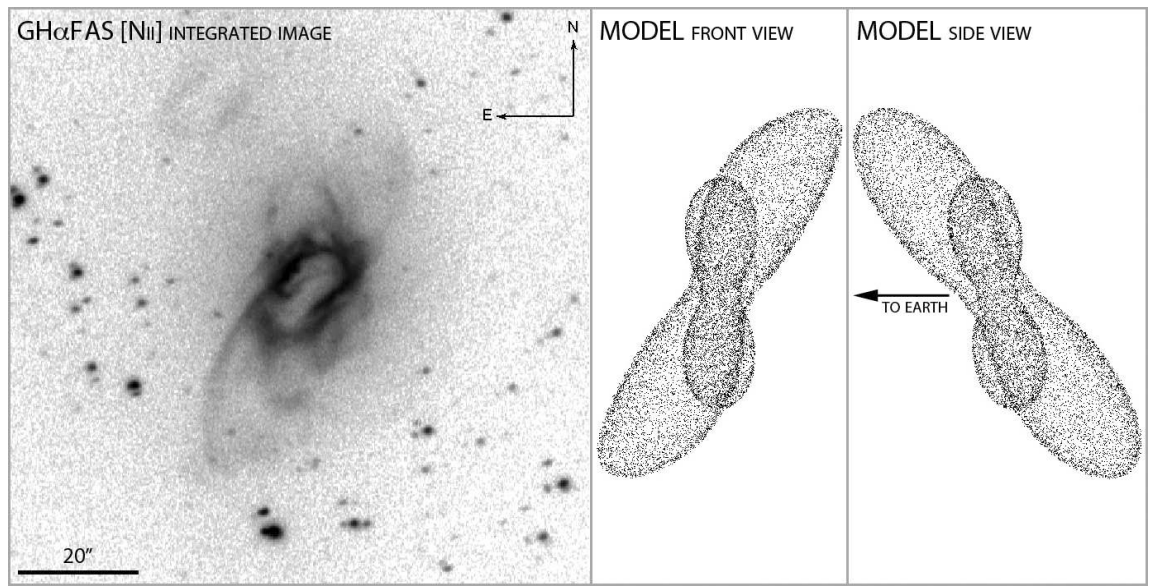}}
\caption{{\bf Left:} GHAFAS datacube integrated across every channel to generate a [N{\sc ii}] image of M~1--75. {\bf Middle:} Adopted model of the small and large lobes (see text) as seen on the plane of the sky. {\bf Right:}  Transversal view of the adopted model.}
\label{F2}
\end{figure*}

The GH$\alpha$FAS [N{\sc ii}] 658.3 nm integrated image of M~1--75 is shown in Fig.~\ref{F2}. While no central star (CSPN) is visible in this image, the nebula clearly shows two pairs of lobes with different orientations. They are nested in a central, brighter rim resembling a horseshoe. Both systems of lobes appear distorted and fragmented, and their faint outer edges are difficult to track near the poles.

The lobes of M~1--75 were the subject of a spatio-kinematical model by \citet{dobrincic08}. From two slit spectra, approximately along each pair of lobes, a [N{\sc ii}] image from \citet{manchado96a}, and simple assumptions such as ballistic expansion, they determined the larger and smaller lobes to lie at inclinations of 87$^{\circ}$ and 65$^{\circ}$, respectively, and to be likely co-eval, with kinematical ages of 2700 and 2400 years per kpc of distance to the nebula, respectively.

Fabry-Perot interferometry (and GH$\alpha$FAS, in particular) represents a significant step forward in spatio-kinematical modelling of planetary nebulae. Not only it allows for a resolution in wavelength comparable to high-resolution echelle spectrographs, but the series of  ``Doppler-map'' images it produces span the whole nebula, instead of being limited by a narrow slit whose orientation has to be decided \emph{a priori} based on previous images. From the spatio-kinematical point of view, a single GH$\alpha$FAS data cube encompasses everything that is needed (i.e. information of the emission both in the plane of the sky and along the line of sight), its quality being only limited by seeing.

In particular, it is noteworthy to remark the faint, high velocity emission regions offset from the axis of the larger lobes (see Fig.~\ref{F3}), near the polar caps (especially in the southwest region). Those certainly would have remained unnoticed, had we been constrained by a narrow slit oriented along the lobes' ``expected'' axis.

\subsection{Solf-Ulrich model}

\begin{figure}[h!]
\center
\resizebox{5cm}{!}{\includegraphics{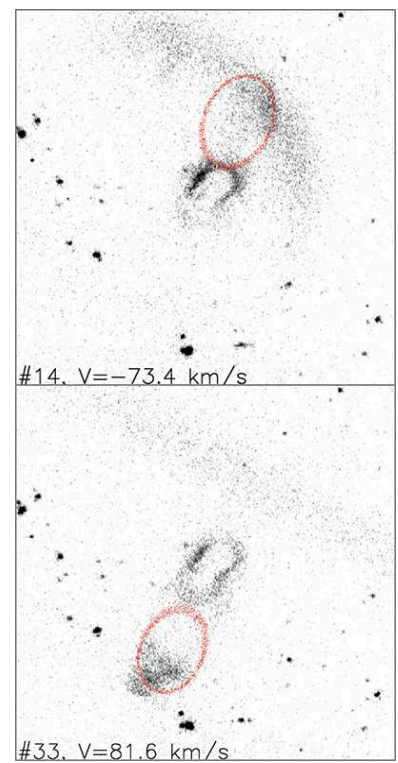}}
\caption{Two GH$\alpha$FAS channels with the initial Solf-Ulrich model of the large lobes overimposed (in red). The $V$ label shows the radial velocity (with respect to the centre of the nebula) of the corresponding channels. The size of each frame is 96$''\times$96$''$. The emission near the poles (specially the southern one) presents a significant offset from the model symmetry axis.}
\label{F3}
\end{figure}

The data cube, with Doppler-shift images spread across 48 channels, allowed us to build a spatio-kinematical model of both systems of lobes (see \citealt{santander04b} for a detailed description of the method). Our first approach for each pair of lobes consisted of fitting a Solf-Ulrich \citep{solf85} surface to the data. This analytical model is described, in spherical coordinates, by:

$$
r = tD^{-1} [v_\mathrm{equator} + (v_\mathrm{polar} - v_\mathrm{equator}) \sin{|\theta|}^\gamma]
$$

where $r$ is the angular distance to the centre of the nebula (i.e. the adopted central star, see section 4),
$tD^{-1}$ the kinematical age of the outflow per unit distance to the nebula, $v_\mathrm{polar}$ and
$v_\mathrm{equator}$ the velocities of the model at the pole and equator
respectively, $\theta$ the latitude angle of the model, and $\gamma$
a dimensionless shaping factor. This assumes that each gas particle travels in the radial direction, with a velocity proportional to its distance to the central source (i.e. in a ``Hubble-like'' flow). 

The two-dimensional generatrix is rotated around the symmetry axis and homogeneously populated with particles to produce a three-dimensional model, and then inclined to the plane of the sky. The resulting geometrical shape and velocity field ---once offset by a certain systemic velocity--- are then used to generate a simplified image of the nebula and a series (one per GH$\alpha$FAS channel) of Fabry-Perot synthetic interferometric images for direct comparison with the [N{\sc ii}] integrated image and GH$\alpha$FAS channels data. The irregular surface brightness distribution is beyond the scope of this paper and therefore has not been fitted.

In order to find the best representation of the nebular geometry and expansion, we allow the parameters to vary over a large range of values and visually compare each resulting model to the integrated image and each of the 48 GH$\alpha$FAS channels, until we obtain the best fit. The range of uncertainty is also derived by eye, by individually changing each parameter away from the best fit, until the resulting model is no longer a fair fit to the data. Note that, although the inclination of each pair of lobes cannot be directly determined from the horseshoe ---a clearly non-elliptical waist---, this fact does not prevent us from finding its value with certain degree of accuracy, given that  the aforementioned parameters are disentangled from one another in the results they produce (i.e. there are no degeneracies in the resulting model).

\begin{figure*}
\center
\resizebox{15cm}{!}{\includegraphics{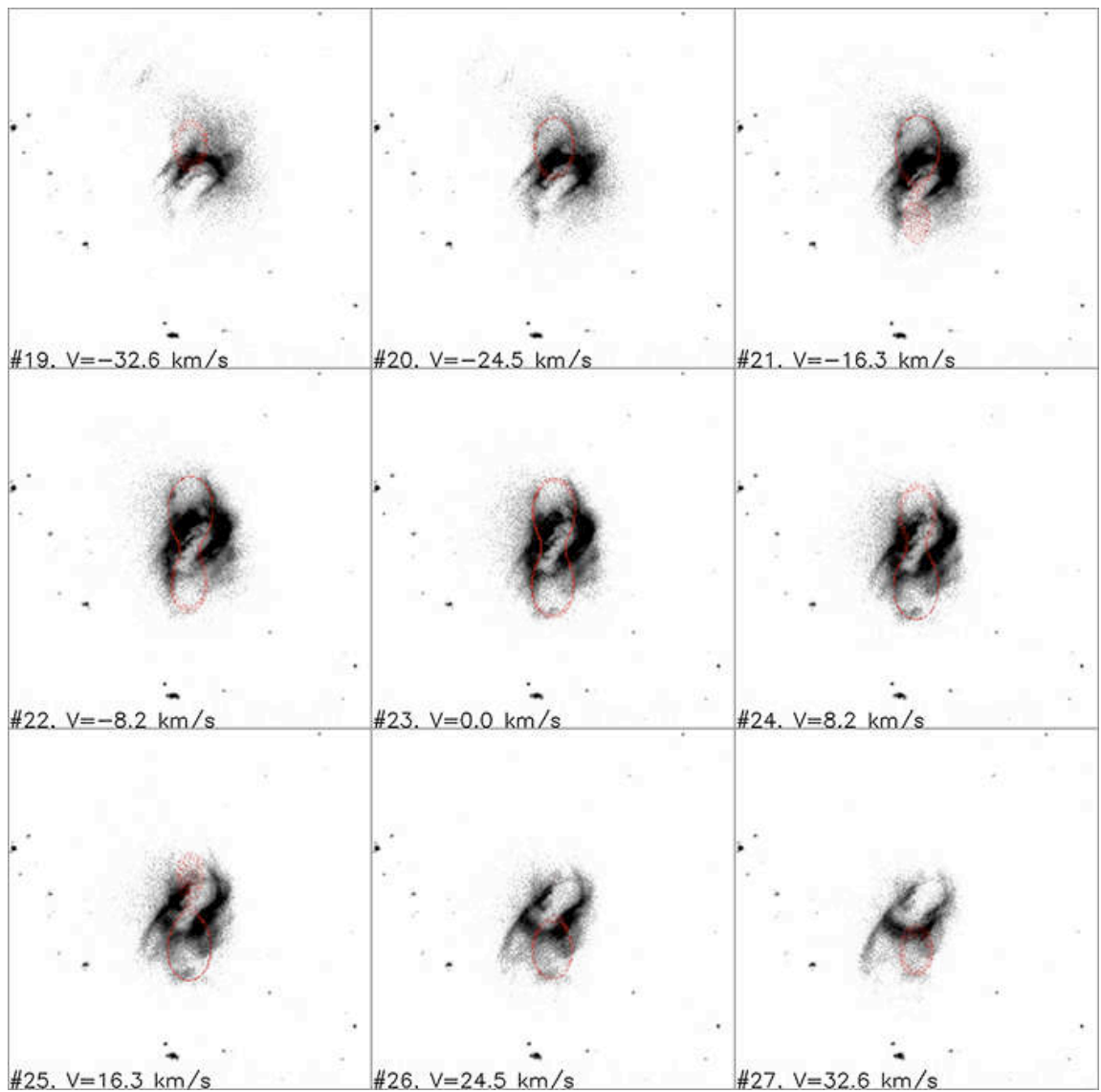}}
\caption{GH$\alpha$FAS data of M~1--75 with the Solf-Ulrich model of the small lobes overimposed (in red). The size of each frame is 96$''\times$96$''$.}
\label{F4}
\end{figure*}

 A fair overall fit to the data was obtained for the large and small lobes (Figs.~\ref{F3} and \ref{F4}), although the model of the former does not account for the offset emission near the poles. The systemic velocity of both system of lobes was found to be 
$v_\mathrm{sys_{LSR}}\sim$13$\pm$4~km~s$^{-1}$. They were also found to share a similar kinematical age of $\sim$1000 yr~kpc$^{-1}$, within uncertainties. The orientations on the sky are instead different ---the larger lobes lie inclined 58$^{\circ}$ to the line of sight, while the inclination of the smaller pair is 79$^{\circ}$. Note that these results (ages, velocities and inclinations) are essentially different from the fit by \citet{dobrincic08}, resulting from two slit positions. We were unable to fit a model with their parameters to the GH$\alpha$FAS data. As their model is based just on two slits rather than the 2-D full kinematical information present in the GH$\alpha$FAS data this is possibly to be expected. The parameters corresponding to our best results for the smaller and larger lobes, together with the uncertainties, are shown in Tables 2 and 3 respectively.

However, no standard Solf-Ulrich model can reproduce the high-velocity emitting region offset from the larger lobes axis. Instead, a modified, point-symmetric  Solf-Ulrich model can account for these structures while still fairly fitting the inner regions.

\subsection{Point-symmetry, modified Solf-Ulrich model}

In order to find a better fit to the GH$\alpha$FAS data, we introduced the following modified Solf-Ulrich model:

$$
r = tD^{-1} [v_\mathrm{equator} + (v_\mathrm{polar} - v_\mathrm{equator}) \sin{|\theta|}^{\gamma(\theta)}] 
$$

where $\gamma(\theta)$ is described by 

$$
\gamma(\theta) = \gamma_\mathrm{equator} + (\gamma_\mathrm{polar}-\gamma_\mathrm{equator}) \ {(\frac{2 \theta}{\pi})}^\epsilon
$$

where $\gamma_\mathrm{equator}$ and $\gamma_\mathrm{polar}$ are the values of the shape factor $\gamma$ at the equator and poles respectively, while $\epsilon$ is the power of the dependance (i.e. 1 linear, 2 quadratic, etc.). This dependance of $\gamma$ with the latitude, although increasing the number of free parameters, allows us to better sample the degree of collimation of the nebula at different latitudes.

The next step was adding point-symmetry to the model. We achieved this in a simple way by defining the nebular axes x, y, z (x along the line of sight towards the viewer, y towards the right, and z upwards along the nebula main axis), and then horizontally projecting the model's z axis on to curves given by

$$
x' = k_x \ z^p
$$

and 

$$
y' = k_y \ z^p
$$

where $k_x$ and $k_y$ are constants, and $p$ is an odd integer (so that it produces a point-symmetric structure). The modified model allows two independent degrees of point symmetry, along the x and y axes, respectively (in a corkscrew fashion). We finally rotated the model by a $\phi$ angle around the z axis before inclining it to the line of sight and produced the synthetic image and GH$\alpha$FAS channel data as described in section 3.1.

Only the large lobes were fit with this model. The best fit values along with their uncertainties ---fully consistent with the standard Solf-Ulrich model except for the curvature--- are listed in the lower part of Table 3.

\begin{figure*}
\center
\resizebox{17cm}{!}{\includegraphics{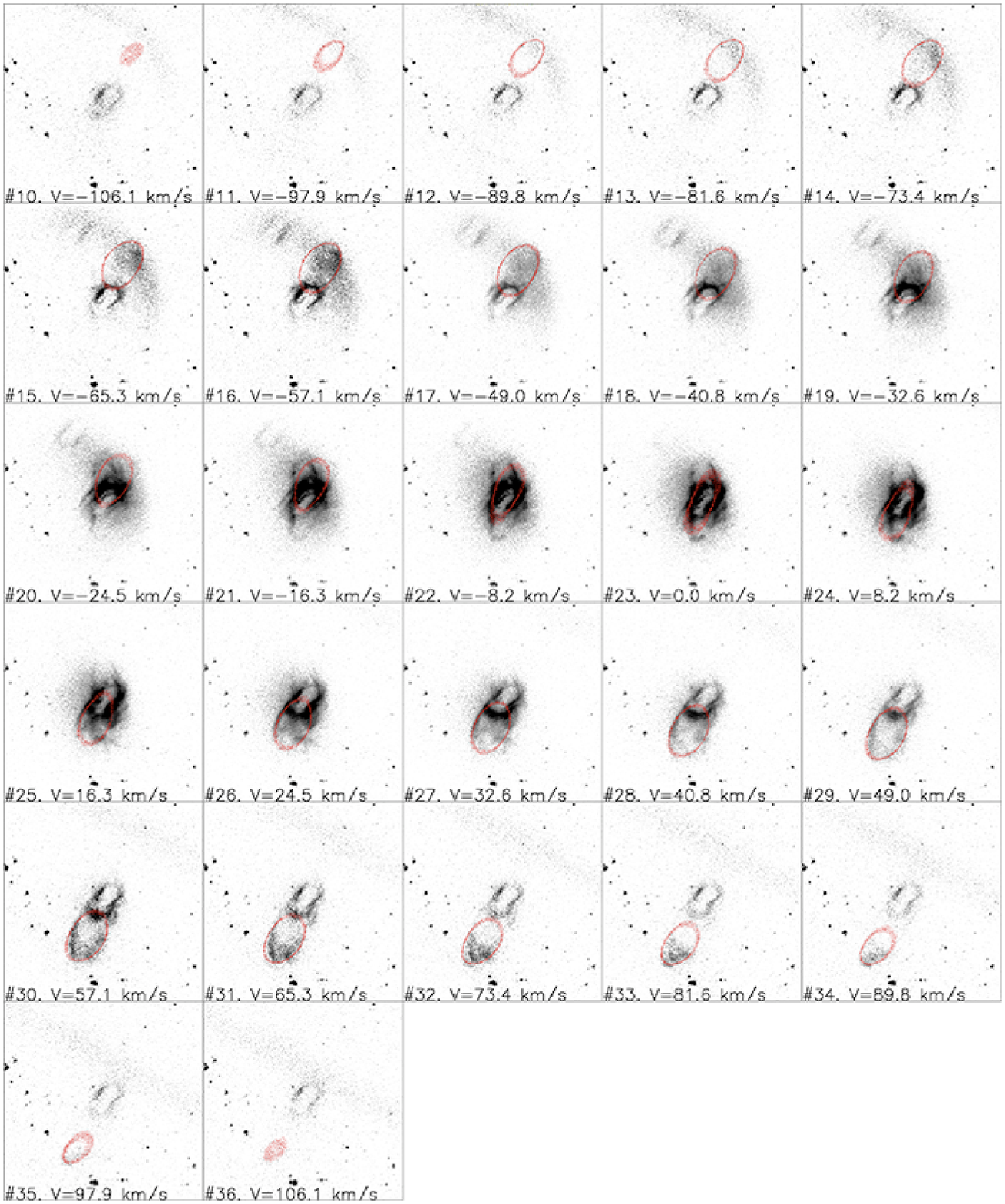}}
\caption{GH$\alpha$FAS data of M~1--75 with the modified Solf-Ulrich, point-symmetric model of the large lobes overimposed (in red). The size of each frame is 96$''\times$96$''$. The brightness/contrast of channels \#17 to \#29 have been modified to provide a better display of the central region, although the noise from the background has also been amplified as a result. The horseshoe ghost is clearly visible near the top left corner in several channels, while the arc-shaped artifact can be seen progressively crossing the frame towards its centre.}
\label{F5}
\end{figure*}

Almost all the emission from the large lobes, including the aforementioned offset region, was found to be faithfully accounted for by the latter model (see Fig.~\ref{F5}), which added a slight corkscrew-like curvature.

\begin{table}[!h] 
\begin{center} 
\begin{tabular}{c c c} 
Parameter & Value & Range\\ 
\noalign{\smallskip} 
\hline\hline 
Small lobes \\
\hline 
\noalign{\smallskip} 
$tD^{-1}$ (yr~kpc$^{-1}$) & 925 & (800-1000)\\
$v_\mathrm{equator}$ (km~s$^{-1}$) & 15-20 & (:)\\
$v_\mathrm{polar}$ (km~s$^{-1}$) & 105 & (90-125)\\
$\gamma$ & 4.5 & (4-5)\\
$P.A.$ ($^\circ$) & 359 & (355-1)\\
$i \ (^\circ)$  & 79 & (76-82)\\
$v_\mathrm{sys_{LSR}}$  &  13 & (11-15)\\
\hline 
\end{tabular} 
\end{center} 
\label{T2} \caption{Best-fitting parameters for the small lobes.  ``:'' means uncertain.} 
\end{table}

\begin{table}[!h] 
\begin{center} 
\begin{tabular}{c c c} 
Parameter & Value & Range\\ 
\noalign{\smallskip} 
\hline\hline
Large lobes \\ 
\hline\hline
Solf-Ulrich model \\ 
\hline 
\noalign{\smallskip} 
$tD^{-1}$ (yr~kpc$^{-1}$) & 1000 & (900-1150)\\
$v_\mathrm{equator}$ (km~s$^{-1}$) & 25 & (23-31)\\
$v_\mathrm{polar}$ (km~s$^{-1}$) & 180 & (160-200)\\
$\gamma$ & 7 & (6.5-7.5)\\
$P.A.$ ($^\circ$) & 337 & (336-339)\\
$i \ (^\circ)$  & 58 & (54-63)\\
$v_\mathrm{sys_{LSR}}$  &  13 & (11-15)\\
\noalign{\smallskip} 
\hline
Point-symmetric model \\ 
\hline 
\noalign{\smallskip} 
$tD^{-1}$ (yr~kpc$^{-1}$) & 1000 & (900-1100)\\
$v_\mathrm{equator}$ (km~s$^{-1}$) & 25 & (23-31)\\
$v_\mathrm{polar}$ (km~s$^{-1}$) & 190 & (170-210)\\
$\gamma_\mathrm{equator}$ & 1 & (1-3)\\
$\gamma_\mathrm{polar}$ & 14 & (13-15)\\
$\epsilon$ & 0.6 & (0.55-0.7)\\
$P.A.$ ($^\circ$) & 338 & (337-339)\\
$i \ (^\circ)$  & 58  & (55-62)\\
$k_x$ & 2$\times$10$^{-5}$ & (1-3)$\times$10$^{-5}$\\
$k_y$ & 8$\times$10$^{-5}$ & (6-9)$\times$10$^{-5}$\\
$p$ & 3 &  - \\
$\phi$ ($^\circ$) & 166 & (150-185)\\
$v_\mathrm{sys_{LSR}}$  &  13 & (11-15)\\
\noalign{\smallskip} 
\hline 
\end{tabular} 
\end{center} 
\label{T3} \caption{{\bf Top:} Best-fitting parameters for the large lobes using a standard Solf-Ulrich model.  {\bf Bottom:} Best-fitting parameters for the large lobes using a point-symmetric, modified Solf-Ulrich model.} 
\end{table}


\section{The central star}

\begin{figure}
\center
\resizebox{8.5cm}{!}{\includegraphics{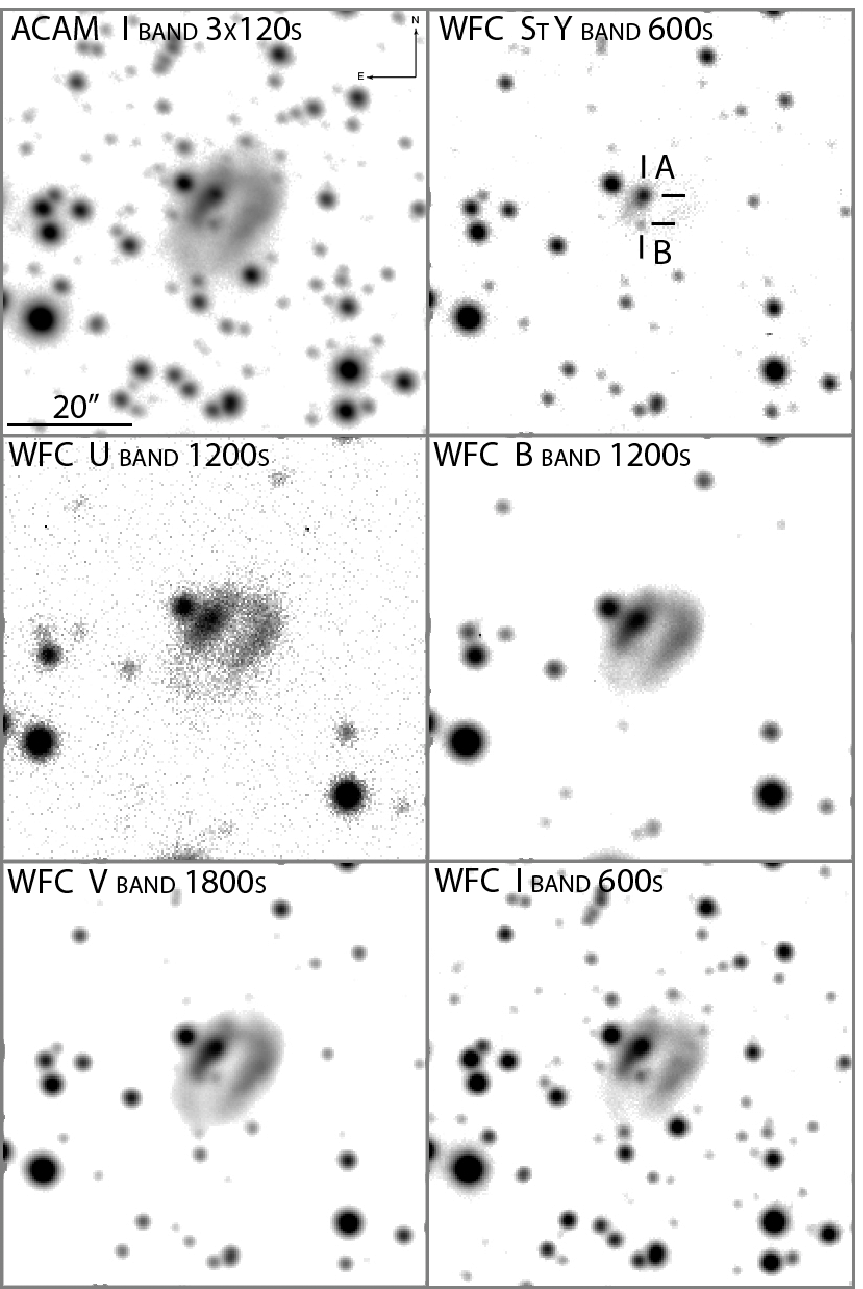}}
\caption{Multiband images of M1~75. Each frame is labelled according to the instrument, band and exposure time. Stars A and B (see text) are labelled in the Strömgren Y image, where the emission from the nebula is practically absent.}
\label{F6}
\end{figure}

The WFC  Str\"omgren Y image (see Fig.~\ref{F6} top right), where the nebular emission is practically absent, shows two faint stars inside the horseshoe region of the nebula. The star labelled as A is offset $\sim$5$''$ with respect to star B, which lies approximately at the centre of the nebular emission. In order to gain some insight on the possibility of either star being the CSPN, we took multi-colour (U, B, V and I) WFC images of the nebula (see Fig.~\ref{F6}) and an ACAM 40~min low-resolution spectrum of both star candidates.

Unfortunately, the spectrum of each star only shows the nebular emission lines together with a continuum whose signal to noise is too low ($\sim$10-15) to allow us to detect any photospheric spectral signatures of a white dwarf. Instead, once the nebular emission has been accounted for, we can estimate the visual magnitudes of both stars. This results in $m_\mathrm{v}\sim19.3$ for star A and $m_\mathrm{v}\sim21.4$ for star B.

On the other hand, star A is barely visible in the images in the light of the U and B bands, and clearly visible in the V and I bands, while star B is only visible in the latter bands. Although those bands are highly contaminated by strong emission from the nebula, we were able to roughly estimate the $V-I$ colour of star B. In order to do this, we added 7 rows ($\sim$1.5$\times$FWHM) centred on the star and with P.A.=138$^\circ$, where the nebular contamination is minimum. For each image, a gaussian was fitted to the star profile, taking a linear fit between the base of the wings of the profile as background. Once the quantum efficiency of the detector (EEV~4280), the filter transmissions and the atmospheric extinction have been taken into account, we found an observed $V-I\sim$2.0 for star B.

The dereddening of this value is not straightforward, due to the extinction variation across the nebula. The extinction values found in the literature range from c$_\mathrm{b}$=2.29 to c$_\mathrm{b}$=2.9 \citep{hua88,bohigas01} from the Balmer decrement for different regions of the nebula. From our own ACAM and IDS spectra, we computed an extinction value c$_\mathrm{b}$=1.9$\pm$0.1 at the location of star B, close to the values found by other authors. If we assume the star to lie within the nebula, we can deredden its $V-I$ colour using the \citet{fitzpatrick04} extinction law, assuming R=3.1, to obtain an intrinsic $(V-I)_\mathrm{0}\sim$0 for star B.


\section{Discussion}

\subsection{The shaping of the nebula}

Our spatio-kinematical modelling confirms the presence of two pairs of nested lobes in the nebula of M~1--75, although with essentially different velocities and ages from those found by \citet{dobrincic08}. For reasons outlined in section 3, we consider our results more reliable. The outer lobes show some evidence of departure from axial symmetry in the polar regions, which we have modelled by applying a slight degree of point-symmetry to an axisymmetric flow. The inner lobes show a different orientation; the spatial angle between their symmetry axes is $\sim$22$^\circ$. The expansion pattern of both systems of lobes is adequately described by a simple Hubble-flow law. In other words, each lobe is the result of a brief, organised shaping process, followed by ballistic expansion. Both systems of lobes share the same age, within uncertainties. 

As the lobes expand, their inner regions would interact and lose their integrity in the process, as shocks progressively convert their kinetic energy to heat. This could explain the broken and essentially irregular structure of the central horseshoe. However, the presence of shocks in the horseshoe is controversial: \citet{guerrero95} and \citet{riera90} found some indications that shock excitation in the horseshoe does not play a significant role in the shaping, while \citet{bohigas01} found proof of shock wave excitation of the H$_\mathrm{2}$ emitting region, which is tightly correlated to that emitting in [N {\sc ii}].  On the other hand, the slight twist at the polar tips of the outer lobes appear to follow the same ballistic expansion pattern as the rest of the structure, thus not arising late in the evolution of the nebula. This could be a clue to the stellar ejection process, perhaps happening in a rapidly rotating frame.

Given all the aforementioned, it is not trivial to depict a formation scenario for M~1--75. The classic Generalised Interacting Stellar Winds (GISW, \citealt{balick87b}, a refined version of the original ISW by \citealt{kwok78}) model. In this model, the isotropic, fast and tenuous wind from a post asymptotic giant branch (AGB) star interacts with the anisotropic, slow and dense winds previously deployed by the star during the AGB stage and shapes a bipolar nebula, is not sufficient to explain either the presence of a multipolar structure, or the slight degree of point-symmetry of the larger lobes. Instead, one has to invoke a mechanism such as the warped-disc proposed by \citet{icke03}: if the post-AGB is surrounded by a disc, warped by radiative instabilities, the wind interaction could result in a multipolar nebula with some degree of point-symmetry in the external regions (e.g. NGC~6537). The origin of the disc itself (i.e. the necessary equatorial density enhancement), however, would still require either the CSPN to be actually a close binary, or to have engulfed one of its planets as it died. A more complex approach is that of \citet{blackman01a}, in which a low-mass companion originates a disc blowing its own wind. A misalignment between the stellar and disc magnetic and rotational axes could give rise to a quadrupolar structure, while the point-symmetry observed in the tips of the larger lobes would require precession of the magnetic axes. We consider the model proposed by \citet{manchado96b}, in which a fast precessing disc is responsible for the different orientation of each structure,  as a less likely scenario, for it would require both structures to have been ejected in extremely rapid succession (in $\sim$300-500 yr to still fit our spatio-kinematical model uncertainties, considering the distance range adopted below and the spatial angular separation of the inner and outer lobes), and it would not explain the point-symmetryic structure.

It is noteworthy to remark that all these models require the CSPN to be/have been a close-binary system (or at least a single star with a close massive hot Jupiter) for the disc to have formed. There is, to our knowledge, no plausible model in the extensive literature able to produce a quadrupolar (not to mention point-symmetric) PN out of a single star.

\subsection{The central star}

Although it constitutes the cornerstone for many parameters of PNe and their central stars, such as the kinematical age of the nebula or their total luminosity, the distance to these objects is poorly known in most cases. M~1--75 is no exception. In the literature one can find several distances based on different methods, such as the statistical distance range, 2.6-3.7~kpc by \citet{cahn92}, or the Galactic rotation curve distance of 5.3~kpc by \citet{burton74}. The more recent extinction-distance method of \citet{sale09}, easily applicable to the INT Photometric H$\alpha$ Survey (IPHAS) data sample and reliable in most PNe \citep{giammanco10}, does not help in the case of nebulae with a significant amount of internal extinction. In the case of M~1--75, the extinction value lies far above the plateau of the field stars in the extinction-distance graph, confirming a significant internal extinction in this nebula  (another possibility would be that stellar H$\alpha$, possibly from a cooler companion, scatters from dust in or near the inner horseshoe, thereby increasing the H$\alpha$/H$\beta$ ratio; this is ruled out, however, by this ratio in the core being lower  than in the horseshoe). Given the lack of evidence favouring a particular distance estimate, rather than adopting a specific distance we will consider a more conservative, intermediate distance range between 3.5 and 5.0~kpc.

Probably due to its internal extinction, so far there has been no clear evidence of the CSPN of M~1--75, other than a slight enhancement of the isophotes in an [O {\sc iii}] image \citep{hua88}. Our images in the light of Str\"omgren Y, followed by low resolution spectra, have detected two candidates to CSPN (Fig.~\ref{F6}), stars A and B (at the position pointed out by Hua), of apparent magnitudes m$_v\sim$19.3 and 21.4 respectively. Unfortunately, the low signal to noise ratio in a 40~min spectrum with ACAM ($\sim$15 in the continuum of the brightest star around 700 nm) prevents us from detecting and analysing any photospheric features, leading us to think that any future research on these stars will require an 8-m class telescope.

Although several nebulae have offset CSPN, it is unlikely that star A is the central star of M~1--75. Even in the most extreme case known (MyCn~18; \citealt{sahai99a}), the star is nowhere in contact with the equatorial waist of its nebula. In fact, to explain such an offset ($\sim$5$''$), one would need to invoke a high proper motion central star travelling $\sim$10-20 km~s$^{-1}$ faster than its own nebula since the ejection, 3500-5000 yr ago at the adopted distance range. The nebula would have to have been heavily braked by interaction with the ISM, being distorted in the process. However, making a simple extrapolation from the PN-ISM interaction models for a round nebula by \citet{wareing07}, every symmetry in the system would have been long lost at such a late stage of interaction.

Therefore we can safely rule out star A and assume star B, at the centre of the nebula, as the CSPN of M~1--75. The visual magnitude we have derived for this star is consistent with the estimate by \citet{hua88} who, assuming a distance of 2.8 kpc \citep{acker78}, suggested a hot (log T$_\mathrm{eff}$=$5.3$) core with a mass of about 0.57-0.6 M$_{\odot}$ with a luminosity of log L/L$_\odot$=2.36. The kinematical age of the nebula found in this work is coherent with the luminosity and T$_\mathrm{eff}$ of the fading evolutionary track of a hydrogen-burning high-mass ($\sim$0.6-0.8 $M_\odot$) core \citep{schonberner93,mendez97}.

Based on the extremely high N/O=2.85 and He/H= 0.18 of the nebula, \citet{guerrero95} hinted towards the possibility of the CSPN actually being a post-common envelope close binary with a total initial mass between 4 and 6 M$_\odot$. In fact, the $(V-I)_0\sim$0 colour estimated in this work is considerably redder than the value of  $V-I$= -0.9 one would expect from a single blackbody of log T$_{eff}$=5.3. This might suggest the presence of  a fainter (L$_\mathrm{bol}<$~10$^{-3}\times$L$_{\mathrm{bol}_\mathrm{WD}}$), much colder (T$_\mathrm{eff}\lesssim$10000 K) companion star, as together they would produce $V-I$ and luminosities coherent with the observations. This would be consistent with the increasing number of confirmed binary cores hosting bipolar PNe (\citealp{miszalski09a}; \citealp{miszalski10}), but would need to be proved via a direct method (e.g. photometric monitoring).


\section{Summary and Conclusions}

A spatio-kinematical model of the M~1--75 nebula has been presented. Two pairs of lobes emerge from the core, their expansion patterns well described by a Hubble-like flow, their kinematical ages ($\sim$1000 yr~kpc$^{-1}$) being similar within uncertainties, while their orientations differ by $\sim$22$^\circ$. The outer lobes have been found to be slightly point-symmetric. The implications of these results on the different shaping theories have been briefly discussed, and a model invoking a close companion star (or a hot Jupiter planet) has been favoured.  

On the other hand, the $V-I$ colour and brightness of the CSPN ---first identified in this work--- are compatible with the presence of a close companion provided its T$_\mathrm{eff}$ is less than 10000~K and its luminosity less than 10$^{-3}$  times that of the white dwarf.

\begin{acknowledgements}
      MSG would like to thank Mariano Santander for his help with the point-symmetric model, and Guillermo Garc\'\i a-Segura for his insight on single stars and quadrupolar nebulae.
\end{acknowledgements}


\bibliographystyle{aa}
\bibliography{msantander}


\end{document}